# Nematic Anchoring on Carbon Nanotubes

Rajratan Basu and Germano S. Iannacchione

*Order-Disorder Phenomena Laboratory, Department of Physics, Worcester Polytechnic Institute, Worcester, MA 01609, USA*

A dilute suspension of carbon nanotubes (CNTs) in a nematic liquid crystal (LC) does not disturb the LC director. Due to a strong LC-CNT anchoring energy and structural symmetry matching, CNT long axis follows the director field, possessing enhanced dielectric anisotropy of the LC media. This strong anchoring energy stabilizes local *pseudo-nematic* domains, resulting in nonzero dielectric anisotropy in the isotropic phase. These anisotropic domains respond to external electric fields and show intrinsic frequency response. The presence of these domains makes the isotropic phase electric field-responsive, giving rise to a large dielectric hysteresis effect.

Nematic liquid crystals (LC) have gained great research interest in recent years for nano-templating purposes, imparting their orientational order onto dispersed nanomaterials [1,2,3,4,5,6,7]. Hybrid liquid crystal-carbon nanotube (CNT) systems increasingly rely on improving electro-optic properties of LCs [4], obtaining directed-self assembly of CNTs over macroscopic dimensions [1,2], and developing nano-electromechanical systems [3]. The CNT alignment mechanism is driven by the coupling of the unperturbed *director field* (average direction of LC molecules) to the anisotropic interfacial tension of the CNTs in the nematic LC matrix, as individual CNTs (not in bundle) are much thinner than the elastic penetration length [8]. So, a dilute suspension is stable because dispersed CNTs, without large agglomerates, does not perturb the director field significantly. Consequently, the suspended nanotubes share their intrinsic properties with the LC matrix, such as electrical conductivity [2], due to the alignment with the LC molecules. A dilute suspension of CNTs in an LC matrix is a unique assemblage of an anisotropic dispersion (CNTs) in an anisotropic media (LC), which makes it an important and active area of research for realizing the LC-CNT interactions and the principles governing CNT-assembly through a nematic mediated platform. We observe that the presence of a small concentration of well-dispersed CNTs in an LC matrix produces enhanced dielectric anisotropy in the nematic phase and nonzero dielectric anisotropy in the isotropic phase. In this letter, we report the ac field-induced dielectric ($\bar{\varepsilon}$) response for multiwall carbon nanotubes (MWCNTs) dispersed in 4-Cyano-4′-Pentylbiphenyl (5CB) LC in both the nematic and isotropic phases.

The nematic phase shows dielectric anisotropy due to the anisotropic nature of the LC molecules where $\varepsilon_\parallel$ and $\varepsilon_\perp$ are the components parallel and perpendicular to the molecular long axis, respectively. For a positive dielectric anisotropic LC, $\varepsilon_\parallel > \varepsilon_\perp$, and so, the director field reorients parallel to an applied electric field. In a uniform homogeneously aligned parallel-plate cell configuration, the nematic director is aligned perpendicular to the applied electric field due to surface anchoring, but the director can reorient parallel to the applied field if the field magnitude is above some critical threshold. This is the essence of a Fréedericksz transition and an ac-capacitive measurement of the $\bar{\varepsilon}$ reveals $\varepsilon_\perp$ below and $\varepsilon_\parallel$ above this switching, the exact values depending on frequency.

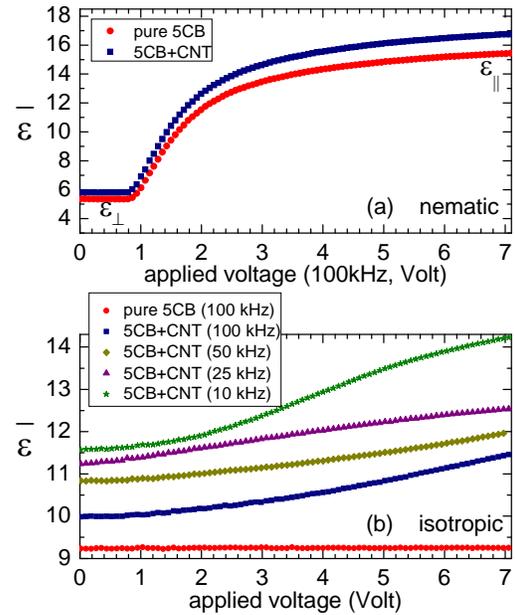

**FIGURE 1:** (a) The average dielectric constant $\bar{\varepsilon}$ as a function of applied ac voltage for 5CB and 5CB+CNT in the nematic phase (T = 23°C); (b) The average dielectric constant $\bar{\varepsilon}$ as a function of applied ac voltage for 5CB and 5CB+CNT in the isotropic phase (T = 45°C).

The hybrid LC-CNT mixture was prepared by dispersing 0.005 wt% of MWCNT sample (containing nanotubes < 8 nm in diameter and 0.5-2 $\mu$m in length) in 5CB (T$_{nematic-isotropic}$ = 35 °C) host via ultrasonication for 5 hours to reach mono-dispersion of CNTs. Soon after ultrasonication, the mixture was degassed under vacuum at 40°C for at least two hours. The mixture then was filled into a homogeneous LC cell (5 × 5 $mm^2$ indium tin oxide (ITO) coated area and 20 $\mu$m spacing [9]) by capillary action. Before performing any dielectric measurements, the CNT doped LC cell was studied under a cross polarized microscope. The micrographs revealed a uniform texture, like a pure LC cell, indicating a uniform nematic director field. There were no indications of phase separation or agglomerates at any temperature. Thus, at least on the length-scales probed by visible light, the structure of

CNT-aggregates (if any) must be small enough that they don't perturb the director field due to their low concentration.

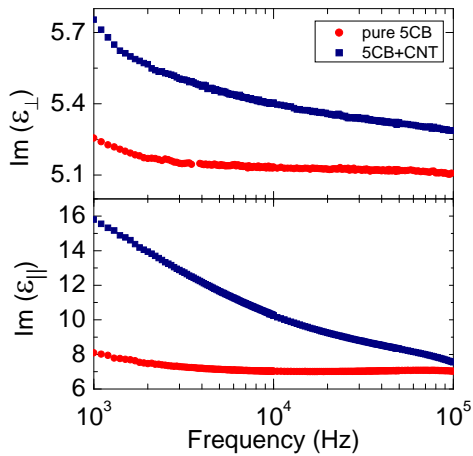

**FIGURE 2:** Imaginary part of $\varepsilon_\perp$ (top panel) and $\varepsilon_\parallel$ (bottom panel) for 5CB and 5CB+MWCNT in the nematic phase (T = 23°C).

The dielectric ($\bar{\varepsilon}$) measurements, as a function of applied ac voltage, frequency, and temperature, were performed by the ac capacitance bridge technique [10,11,12] in the nematic and isotropic phases. The reason for applying the ac voltage (not dc) is to avoid the affect of ion migration on the dielectric measurements. The LC 5CB does not exhibit any tumbling mode [13] and MWCNTs show no space charge or dipole orientation dynamics [14] at the probing frequencies. Thus, the observed increase in $\bar{\varepsilon}$ as a function of applied ac voltage should be driven mainly by a mechanical rotation of the nematic director field.

Figure 1 (a) shows the average dielectric constant ($\bar{\varepsilon}$) as a function of applied ac voltage for 5CB and 5CB+MWCNT sample in the nematic phase (T = 23° C). As seen in the Fig. 1, both the pure LC and the mixture undergo planner ($\varepsilon_\perp$) to homeotropic ($\varepsilon_\parallel$) orientational transition, starting at around 0.9 V, saturating at around 7 V. The dielectric anisotropy ($\Delta\varepsilon = \varepsilon_\parallel - \varepsilon_\perp$) for pure 5CB is around +10 [15], as also observed experimentally ($\Delta\varepsilon_{LC} = +10.1$) in Fig. 1(a). The presence of CNTs in LC results in an increase in the dielectric anisotropy ($\Delta\varepsilon_{LC+CNT} = +11.1$), enhancing the nematic order in the nematic matrix. It is important to point out that this dramatic increase occurred due to the addition of only 0.005 wt% MWCNT sample. It has been experimentally shown that the isotropic to nematic phase transition temperature of a liquid crystal is enhanced by the incorporation of small amount of MWNT sample [16], indicating an improvement in the nematic order. A recent study shows that CNTs induce alignment on the nematic LC director field along their long axes due to LC-CNT anchoring effect [6]. Theoretical calculations predict that the strong interaction associated with the CNT alignment mechanism is mainly due to surface anchoring with a binding energy $U_{anchor} = -2$ eV for $\pi$–$\pi$ stacking between CNT and LC molecules [4,17]. As the nanotubes used for this work are not ferroelectric in nature, the additional ordering effect must not be due to the electronic coupling of any permanent dipole moments with the LC dielectric anisotropy; which generally occurs in ferroelectric nanoparticle suspensions in the LCs [18]. Thus, the increase in $\Delta\varepsilon$ (hence, improvement in nematic order) is attributed to the anchoring energy and anisotropic structure of CNTs. The reorientation threshold voltage $V_{th}$ can be estimated by the expression for the Fréedericksz transition [19]: $V_{th} \propto \sqrt{K/\Delta\varepsilon}$; where $K$ is the elastic constant for bent distortion. From Fig 1 (a), as $V_{th}^{LC} \simeq V_{th}^{LC+CNT}$, one can write, $\sqrt{K_{LC}/\Delta\varepsilon_{LC}} \simeq \sqrt{K_{LC+CNT}/\Delta\varepsilon_{LC+CNT}}$; which gives $K_{LC+CNT} \simeq 1.21 K_{LC}$. The strong elastic interaction between CNTs and LC molecules due to surface anchoring may increase the elastic energy of the hybrid system and therefore can be attributed to the increase in $K$ in the LC+CNT system.

As the dielectric constant of a material is determined by the structural arrangement and molecular polarizability, the dielectric spectra for 5CB and 5CB+CNT have been studied in the range $10^3 – 10^5$ Hz to probe any structural modification in hybrid 5CB+CNT system. The spectra, shown in Fig. 2, reveal significant difference in dielectric behavior between 5CB and 5CB+CNT, confirming a structural modification in the nematic phase due to the addition of a small amount of CNT sample.

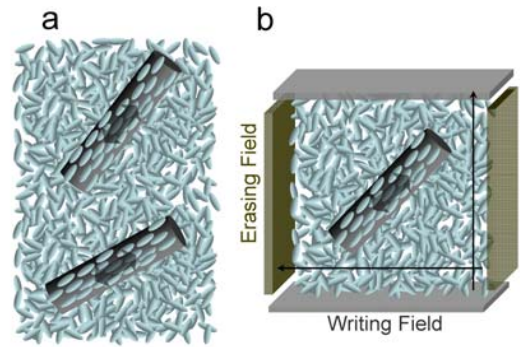

**FIGURE 3:** Schematic diagrams; (a) presence of field-responsive anisotropic *pseudo-nematic* domains due to LC (ellipsoidal) –CNT (cylindrical) interaction in the isotropic media; (b) a model for a four-electrode LC cell for writing and erasing memory.

Due to the absence of elastic interactions in the isotropic phase, the LC molecules no longer maintain long range orientation order and act as an isotropic liquid. Isotropic phase of 5CB, as expected, does not respond to an external field, as also experimentally confirmed in Fig. 1(b). But, a dramatic change in the field-induced dielectric constant has been observed in the isotropic phase for the LC+CNT system. The composite system shows an increment in $\bar{\varepsilon}$ with increasing applied voltage, as shown in Fig 1(b). As stated earlier, the energy associated with LC-CNT anchoring mechanisms is $|U_{anchor}| = 2$ eV, which is much more than the thermal energy, $U_{thermal} \sim k_B T = 2.74 \times 10^{-2}$ eV, for the deep isotropic phase at T = 45°C = 318 K. So, the thermal energy is not even close enough to eliminate the anchoring mechanisms in the deep isotropic phase. Due to this surface anchoring, the CNT induces local short-range orientation order of LC molecules surrounding the CNT having local director along the tube axis, which can be visualized as presence of isolated *pseudo-nematic* domains in an isotropic media, as schematically shown in Fig. 3(a). As these local anisotropic *pseudo-nematic* domains have polarization, their *short-range director field* interact with external electric fields – hence, observed increase in $\bar{\varepsilon}$ with increasing applied voltage in the isotropic phase. These nanoscale anisotropic domains seem to have strong frequency response as the change in $\bar{\varepsilon}$ is different for different frequencies from 10 to 100 kHz in the same applied voltage range, observed in Fig. 1(b).

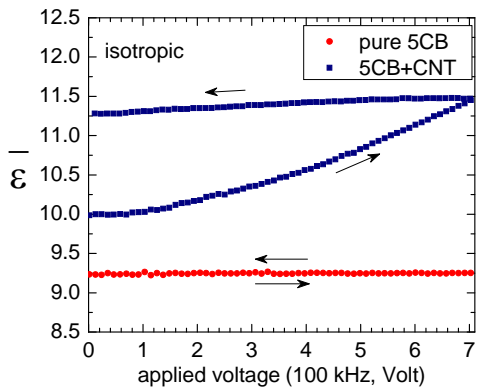

**FIGURE 4:** Dielectric hysteresis for 5CB and 5CB+CNT. The arrows show the cycling direction of the applied ac voltage.

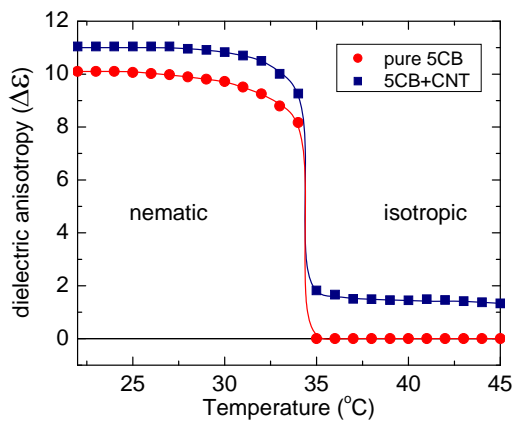

**FIGURE 5:** Dielectric anisotropy as a function of temperature for 5CB and 5CB+CNT. Lines represent guide to the eye.

Dielectric hysteresis has been studied in the isotropic phase (T = 45°C) to understand the stability of these *pseudo-nematic* domains. See Fig. 4. The isotropic phase for pure 5CB does not show any hysteresis effect, as expected, as it does not respond to an external electric field. After reaching the maximum dielectric value at the highest applied field, $\bar{\varepsilon}$ for the composite system does not relax back to its original value cycling the field down to zero, showing a large dielectric hysteresis unlike the nematic phase. As there is no long range orientational order, there is no elastic interaction present in the isotropic LC media. As a result, when the field goes off, there is no restoring force to mechanically torque these domains back into the original orientation in the isotropic phase and the domains stay oriented. This indicates that CNTs, surrounded by few layers of LCs, stabilize nematic-like short-range order in the isotropic phase, giving rise to a ferroelectric type hysteresis. This is the essence of a non-volatile electro-mechanical memory effect. This memory may be erased by applying another field in the opposite direction of the first one in a four-electrode cell configuration as shown in Fig. 3(b). In that case the writing and erasing time can be in the order of the tumbling relaxation mode of the LC, which is 0.1 GHz [13] for 5CB. However, the time response would also depend on the distribution symmetry of the CNTs in the LC media.

The extracted dielectric anisotropy Δε for 5CB and 5CB+CNT is plotted as a function of temperature in Fig. 5. The composite system shows enhanced Δε in the nematic phase. As the order parameter [19] for the pure isotropic phase is zero, Δε for pure 5CB drops down to zero in the isotropic phase. The composite system shows a nonzero value of Δε [20], indicating the presence a net residual order in the isotropic phase. The value of Δε in the isotropic phase changes with increasing frequency in the frequency-range (10 – 100 kHz) studied.

In conclusion, we have observed that dispersing a low concentration of CNTs in a nematic LC, results in an improvement in the nematic ordering, showing enhanced Δε. The presence of local anisotropic *pseudo-nematic* domains in the isotropic phase in the system causes ferroelectric-type hysteresis effects which could find potential applications in memory functions. Future work involves simultaneous dielectric and optical studies on size dependent CNT suspensions in LC media to investigate more on the anchoring effect in both the nematic and isotropic phases.